\documentclass[prd,aps,preprintnumbers,floats,floatfix,superscriptaddress,preprintnumbers,
showpacs,eqsecnum,nofootinbib,twocolumn]{revtex4-1}
\usepackage[utf8]{inputenc}
\usepackage{psfrag}
\usepackage{amsfonts,amsmath,amssymb,latexsym,array,afterpage,theorem,mathrsfs,bm,float,tabularx,here,color,graphicx,multirow}

\newcommand{\nn}{\nonumber \\}
\newcommand{\bea}{\begin{eqnarray}}
\newcommand{\ena}{\end{eqnarray}}
\newcommand{\beann}{\begin{eqnarray*}}
\newcommand{\enann}{\end{eqnarray*}}

\newcommand{\gsim}{\, \mbox{\raisebox{-1.ex}
{$\stackrel{\textstyle>}{\textstyle\sim}$}}\,}
\newcommand{\lsim}{\, \mbox{\raisebox{-1.ex}
{$\stackrel{\textstyle<}{\textstyle\sim}$}}\,}

\newcommand{\vect}[1]{\!\!\!\mbox{~\,\boldmath $#1$}}

\begin{document}
\baselineskip=12pt

\preprint{WU-AP/1803/18}

\title{Maximal Efficiency of Collisional Penrose Process with Spinning Particles}

\author{Kei-ichi Maeda}
\email{maeda"at"waseda.jp}
\affiliation{
Department of Physics, Waseda University,
Shinjuku, Tokyo 169-8555, Japan}
\author{Kazumasa Okabayashi}
\email{bayashioka"at"gravity.phys.waseda.ac.jp}
\affiliation{
Department of Physics, Waseda University,
Shinjuku, Tokyo 169-8555, Japan}
\author{Hirotada Okawa}
\email{h.okawa"at"aoni.waseda.jp}
\affiliation{Yukawa Institute for Theoretical Physics, Kyoto University, Kyoto 606-8502, Japan}
\affiliation{Advanced Research Institute for Science and Engineering, Waseda University, Tokyo 169-8555, Japan}

\date{\today}

\begin{abstract}
We analyze collisional Penrose process of spinning test particles in an extreme Kerr black hole. We consider that two  particles plunge into the black hole from infinity and collide near the black hole. 
For the collision of two massive particles, if the spins 
of particles are $s_1\approx 0.01379\mu M$ and  $s_2\approx -0.2709\mu M$,
we obtain the maximal efficiency is about $\eta_{\rm max}=({\rm extracted~energy })/ ({\rm input~energy})\approx 15.01$, which is 
more than twice as large as the case of the collision of non-spinning particles ($\eta_{\rm max}\approx 6.32$).
We also evaluate the collision of a massless particle without spin and a massive particle with spin (Compton scattering),  in which we find the maximal efficiency is 
$\eta_{\rm max}\approx 26.85$ when $s_2\approx -0.2709\mu M$, which should be compared with $\eta_{\rm max}\approx 13.93$  for the nonspinning case.

\end{abstract}

\maketitle




\section{Introduction} 
A black hole is the most strongly bound system. 
If we can extract energy from a black hole, it would be much more efficient than nuclear energy.
However, because of 
the black hole area theorem\cite{Hawking:1971}, we cannot extract energy 
from a Schwarzschild black hole.
For a rotating black hole, instead, Penrose 
suggested the use of the ergoregion of a rotating black hole
to extract energy\cite{Penrose:1969pc}.
 A particle can have negative energy in the ergoregion.
Hence we suppose that  a plunged particle in the ergoregion 
breaks up into two particles such that  
one particle has negative energy and 
 falls into the black hole,  while
the other particle with positive energy,
which is larger than the  input energy, 
goes away to infinity.
As a result, we can extract energy from a rotating black hole, 
which is called Penrose process.

It was pointed out that this Penrose process 
could play a key role in the energy emission 
mechanism of jets and/or X-rays from astrophysical objects
\cite{Wheeler:1970}.
It has become one of the most interesting and important
mechanisms in astrophysics as well as in general relativity.
However, 
some earlier works
\cite{Bardeen:1972fi,Wald:1974kya,kovetz1975efficiency}
showed that 
the incident particle or the break-up particles
must be relativistic, which implies that
the Penrose process is rare in astrophysics and that 
this process cannot serve for astrophysical process. 

A disintegration of a plunged particle may also not be practical for  extraction of energy from a black hole. Hence two more plausible methods have been proposed: One is a  superradiance, in which we use propagating waves instead of a particle \cite{Zeldovich:1971, Zeldovich:1972, Misner:1972, Press:1972, Bekenstein:1973}.
An impinging wave on a rotating black hole is amplified for some range of frequencies when it is
scattered (see \cite{Brito:2015oca} for the recent progress).
The other one is a collisional Penrose process, in which 
two particles plunge into a black hole and collide in the ergoregion
instead of disintegration of a single particle\cite{Piran1975}.
One expects that it may 
give more efficient mechanism in astrophysical situations.
Unfortunately, the efficiency of the energy extraction,
which is the ratio of the extracted energy 
to the input energy, 
turns to be as modest as the original Penrose process
\cite{piran1977upper}.

Recently this process has again attracted much attention
because Ba\~nados, Silk and West\cite{Banados:2009pr} 
showed that the center of mass energy of two particles 
can be arbitrarily large
when the angular momentum of one incident particle is tuned
and the collision occurs near the horizon of an extreme Kerr black hole.
This is referred to the BSW effect.
If the center-of-mass energy is enough large, new unknown particles 
could be created if any.
It may reveal new physics. 
It could also play an  important role in astrophysics.

There have been so far many studies on the BSW effect after their finding\cite{Jacobson:2009zg,Berti:2009bk,PhysRevD.82.103005,Banados:2010kn,Zaslavskii:2010jd,Zaslavskii:2010aw,Grib:2010dz,Lake:2010bq,Harada:2010yv,Kimura:2010qy,Patil:2011yb,Abdujabbarov:2013qka,Tsukamoto:2013dna,Toshmatov:2014qja,Tsukamoto:2014swa,Armaza:2015eha}.
Since the interaction between a black hole spin and an angular momentum of the particle is essential
for the Penrose process and the BSW effect, it may be interesting to discuss 
collision of spinning particles.
As we will summarize in the text, the 4-momentum of a spinning particle 
is not always parallel to its 4-velocity, resulting  in the possibility of 
violation of the timelike condition of the orbit.
As a result,  although the BSW effect by collision of spinning particles in nonrotating 
Schwarzschild spacetime can take place near the horizon,  the motion of the spinning particles 
becomes superluminal before the collision point\cite{Armaza:2015eha}.
While, if the particle energy satisfies $E<\sqrt{3}\mu /6 $,
with which such a particle cannot plunge from infinity,
the timelike condition is preserved until the horizon\cite{Zaslavskii:2016dfh}.
Of course, we find the BSW effect for the collision of spinning particles in a rapidly rotating Kerr (or Kerr-Newman) black hole\cite{Guo:2016vbt, Zhang:2016btg}.

We are also very curious about the efficiency of energy extraction 
from a black hole, which is defined by $\eta=$(output energy)/(input energy).
Even when the center-of-mass energy becomes  arbitrarily large near the horizon,
a resulting particle may not necessarily 
escape to infinity. 
Thus, it is also important to study how large is the efficiency of the energy extraction from a black hole.

When two massive particles collide near the horizon  on the equatorial plane and 
are converted  to massless particles (photons), 
Bejger et al\cite{Bejger:2012yb}
 showed  numerically that the maximal efficiency is about 1.29.
This result has been confirmed analytically by  Harada, Nemoto and Miyamoto\cite{Harada:2012ap}. However, as 
Schnittman showed numerically\cite{Schnittman:2014zsa},
 the maximal efficiency becomes 13.92 
when an outgoing fine-tuned  massless particle collides 
with a massive particle near the horizon. 
Leiderschneider and Piran\cite{Leiderschneider:2015kwa}
then derived  the maximal efficiency analytically for several possible processes.
They analyzed not only the collision on the equatorial plane 
but also more general off-plane orbits. 
They concluded that  the maximal efficiency is $(2+\sqrt{3})^2\approx 13.93$,
which is found in the case of the Compton scattering
(collision of massless and massive particles)
on the equatorial plane. 
The similar analytic approaches were performed in \cite{Ogasawara:2015umo} and \cite{Zaslavskii:2016unn}.
These results agree with the numerical result by Schnittman\cite{Schnittman:2014zsa}

More efficient way of extracting the energy from a black hole, 
which is called the super-Penrose process, 
has been proposed in  \cite{Berti:2014lva,Patil:2015fua}, 
but there is still an argument \cite{Leiderschneider:2015kwa}.
The essential problem is how to create the particles which cause the super-Penrose process.
Zaslavskii\cite{Zaslavskii:2015fqy} pointed out that it is difficult to
prepare a suitable initial state only by preceding mechanical collisions.

One natural question may arise: How the efficiency of the collisional Penrose process 
will be enhanced when the particles are spinning?
Recently this subject was discussed in \cite{MUKHERJEE201854}.
However the timelike condition was not properly taken into account.
The value of spin is too large for the orbit to be timelike.
Here we will study the effect of the particle spin on the efficiency of energy extraction
in detail. 
We consider the  collision of two massive spinning particles
and the Compton or inverse Compton scattering
 (collision of one massless and one massive particles).
In Sec. II, we briefly review the equation of motion of a spinning particle in a Kerr black hole 
and provide the timelike condition of the orbit.
In Sec. III, we study the collision of two spinning particles
 in an extreme Kerr geometry and 
analyze the maximal efficiency.
We also discuss the collision of one spinning massive particle and one massless particle
( the Compton and the inverse Compton scatterings) .
Section IV is devoted to concluding remarks.
Throughout this paper,
we use the geometrical units of $c=G=1$ and follow \cite{misner1971gravitation}
for the notations.
\\[2em]

\section{Basic Equations}
\label{basic_equations}

\subsection{Equations of Motion of a Spinning Particle}
We consider a spinning particle in Kerr geometry.
The equations of motion of a spinning particle were first derived
by Papapetrou\cite{papapetrou1951spinning} 
by the use of  the pole-dipole approximation of an extended body, 
and then reformulated by Dixon\cite{dixon1970:1,dixon1970:2,dixon1979isolated}.
The equations of motion are
\beann
&&
{Dp^\mu\over d\tau}=-{1\over 2} R^\mu_{~\nu\rho\sigma} v^\nu S^{\rho\sigma}\\
&&
{DS^{\mu\nu}\over d\tau}=p^\mu v^\nu-p^\nu v^\mu
\enann
where $p^\mu, v^\mu=dz^\mu/d\tau, $ and $S^{\mu\nu}$ are the 4-momentum, the 4-velocity and the spin tensor of the particle, respectively. $\tau$ is the proper time and $z^\mu(\tau)$ is the orbit of the particle. 
We need a set of supplementary conditions
\beann
S^{\mu\nu}p_\nu=0
\,,
\enann
which fixes the center of mass of the particle.

 Defining the particle mass $\mu (>0)$ 
by $\mu^2=-p^\mu p_\mu$, 
we also use a specific 4-momentum $u^\mu$,
which is defined by 
\beann
u^\mu={p^\mu\over \mu}\,.
\enann
The normalized magnitude of spin $s$ is defined by 
\beann
S^{\mu\nu}S_{\mu\nu}=2\mu^2 s^2
\,.
\enann
We also  normalize the affine parameter $\tau$ 
as
\beann
u^\mu v_\mu=-1\,.
\enann
We then find the relation between the 4-velocity and the specific 4-momentum as
\beann
v^\mu-u^\mu={S^{\mu\nu}R_{\nu\rho\sigma\lambda}u^\rho S^{\sigma\lambda}
\over 2(\mu^2+{1\over 4}R_{\alpha \beta\gamma\delta}S^{\alpha \beta}S^{\gamma\delta})}
\,,
\enann
which means that the 4-velocity $v^\mu$ and the 4-momentum $p^\mu$ are not always parallel.

\subsection{Conserved Quantities}

If we have a Killing vector $\xi_\mu$ in a background geometry,
 we obtain the conserved quantity
 \beann
 Q_\xi=p^\mu\xi_\mu+{1\over 2}S^{\mu\nu}\nabla_\mu \xi_\nu
 \,.
 \enann

 In the Kerr geometry, there are two Killing vectors:
 \beann
&&
 \xi_{\mu}^{(t)}=-\left(\sqrt{\Delta\over \Sigma}e_\mu^{(0)}+{a\sin \theta\over \sqrt{\Sigma}}e_\mu^{(3)}\right)
 \\
 &&
\xi_{\mu}^{(\phi)}=a\sqrt{\Delta\over \Sigma}\sin^2\theta e_\mu^{(0)}
 +{(r^2+a^2)\sin \theta\over \sqrt{\Sigma}}e_\mu^{(3)}
 \,,
 \enann
 where
 \beann
 \Delta&=& r^2-2Mr+a^2
 \\ 
 \Sigma&=& r^2+a^2\cos^2\theta
 \,,
 \enann
and the tetrad basis $e_\mu^{~(a)} $  is defined by
\[
e_\mu^{~(a)}  = \left(
    \begin{array}{cccc}
      \sqrt{\Delta\over \Sigma} & 0 & 0& -a\sqrt{\Delta\over \Sigma}\sin^2\theta\\
      0 &\sqrt{\Sigma\over \Delta} & 0& 0\\
     0& 0 & \sqrt{\Sigma}& 0\\
    -{a\over \sqrt{\Sigma}} \sin\theta & 0 & 0&{(r^2+a^2) \over \sqrt{\Sigma}}\sin\theta
    \end{array}
  \right)
\,.
\]

Hence there are two conserved quantities in Kerr geometry, which are
the energy $E$ and the $z$ component of the total angular momentum $J$
given by
\beann
E&:= &-Q_{\xi^{(t)}}
\\
&=&\sqrt{\Delta\over \Sigma} p^{(0)}+{a\sin\theta\over \sqrt{\Sigma}}p^{(3)}
\\
~~&&
+{M(r^2-a^2\cos^2\theta)\over \Sigma^2}S^{(1)(0)}+
{2Mar\cos\theta\over \Sigma^2}S^{(2)(3)}
\\
J&:= &Q_{\xi^{(\phi)}}
\\
&=&a\sin^2\theta \sqrt{\Delta\over \Sigma} p^{(0)}+{(r^2+a^2)\sin\theta\over \sqrt{\Sigma}}p^{(3)}
\\
~~&&
+{a\sin^2\theta \over \Sigma^2}[(r-M)\Sigma+2Mr^2]S^{(1)(0)}
\\
~~&&
+{a\sqrt{\Delta}\sin\theta \cos\theta\over \Sigma}S^{(2)(0)}
+{r\sqrt{\Delta}\sin\theta \over \Sigma}S^{(1)(3)}
\\
~~&&+
{\cos\theta\over \Sigma^2}[(r^2+a^2)^2-a^2\Delta\sin^2\theta]S^{(2)(3)}
\,.
\enann
 
\subsection{Equations of Motion in the Equatorial Plane}
We introduce a specific spin vector $s^{(a)}$ by
\beann
s^{(a)}=-{1\over 2\mu}\epsilon^{(a)}_{~(b)(c)(d)}u^{(b)}S^{(c)(d)}
\,,
\enann
which is inversed as
\beann
S^{(a)(b)}=\mu\epsilon^{(a)(b)}_{~~~~(c)(d)}u^{(c)}s^{(d)}
\,,
\enann
where $\epsilon_{(a)(b)(c)(d)}$  is the totally antisymmetric tensor with 
$\epsilon_{(0)(1)(2)(3)}=1$.

In what follows, we consider only the particle motion in the equatorial plane ($\theta=\pi/2$)\cite{Saijo:1998mn}.
From this constraint, we find that the spin direction is always perpendicular to the 
equatorial plane.
Hence only one component of $s^{(a)}$ is nontrivial, i.e., 
\beann
s^{(2)}=-s
\,.
\enann
If $s>0$, the particle spin is parallel to the black hole
 rotation, while when $s<0$, it is antiparallel.

As a result, the spin tensor is described as
\beann
S^{(0)(1)}=- s p^{(3)}\,,~~S^{(0)(3)}= s p^{(1)}\,,~~S^{(1)(3)}= s p^{(0)}\,.
\enann
We then obtain the conserved quantities as
\beann
E&=&{\sqrt{\Delta}\over r}p^{(0)}+{(ar+Ms)\over r^2}p^{(3)}
\label{Eu0u3}
\\
J&=&
{\sqrt{\Delta}\over r}(a+s)p^{(0)}+{r(r^2+a^2)+as (r+M) \over r^2}p^{(3)}.
~~~~~~~~
\label{Ju0u3}
\enann

From those equations, we find
\beann
u^{(0)}&=&
{\left[
(r^3+a(a+s)r+aMs)E-(ar+Ms)J\right]\over \mu r^2\sqrt{\Delta}\left(1-{Ms^2\over r^3}\right)}~~~~~~~~~~
\label{u0EJ}
\\
u^{(3)}&=&
{\left[
J-(a+s)E\right]\over \mu r\left(1-{Ms^2\over r^3}\right)}
\,.
\label{u3EJ}
\enann

There exists the normalization condition $u_\mu u^{\mu}=-1$, i.e., 
\beann
-(u^{(0)})^2+(u^{(1)})^2+(u^{(3)})^2=-1
\,.
\enann
Hence we have 
\beann
u^{(1)}&=&\sigma \sqrt{(u^{(0)})^2-(u^{(3)})^2-1}\,,
\enann
where $\sigma=\pm 1$ correspond to the outgoing and ingoing motions, respectively.

The relation between the 4-velocity $v^{(a)}$ and the specific 4-momentum 
$u^{(a)}$ is given by
\beann
v^{(0)}&=&\Lambda_s^{-1} u^{(0)}\,,~~
\\
v^{(1)}&=&\Lambda_s^{-1}  u^{(1)}\,,~~
\\
v^{(3)}&=&{\left(1+{2Ms^2\over r^3}\right)\over 
\left(1-{Ms^2\over r^3}\right)} \Lambda_s^{-1} u^{(3)}\,,
\enann
where
\beann
\Sigma_s&=&r^2\left(1-{Ms^2\over r^3}\right)
\\
\Lambda_s&=&1-{3Ms^2r[J-(a+s)E]^2\over \mu^2 \Sigma_s^3}
\,.
\enann

Hence 
we obtain
\beann
{dt\over d\tau}&:=&v^{0}= {r^2+a^2\over r\sqrt{\Delta}}v^{(0)}
+{a\over r} v^{(3)}
\\&&={1\over r \Lambda_s}\left({r^2+a^2\over \sqrt{\Delta}}u^{(0)} +a{1+{2Ms^2\over r^3}\over 1-{Ms^2\over r^3}}
u^{(3)}\right)\,,~~
\\
{dr\over d\tau}&:=&v^{1}=  {\sqrt{\Delta}\over r}v^{(1)}=  {\sqrt{\Delta}\over r\Lambda_s }u^{(1)}\,,~~
\\
{d\phi\over d\tau}&:=&v^{3}= {a\over r\sqrt{\Delta}}v^{(0)}
+{1\over r} v^{(3)}
\\&&={1\over r \Lambda_s}\left({a\over \sqrt{\Delta}}u^{(0)} +{1+{2Ms^2\over r^3}\over 1-{Ms^2\over r^3}}
u^{(3)}\right)\, .
\enann

\begin{widetext}
We finally obtain the equations of motion of the spinning particle  as
\beann
\Sigma_s \Lambda_s \mu {dt\over d\tau}&=& 
{\Sigma_s  \mu \over r }\left({r^2+a^2\over 
\sqrt{\Delta}}u^{(0)} +a{1+{2Ms^2\over r^3}\over 1-{Ms^2\over r^3}}
u^{(3)}\right)= a\left(1+{3Ms^2\over r\Sigma_s}\right)[
J-(a+s)E]+{r^2+a^2\over \Delta}P_s
\\
\Sigma_s \Lambda_s \mu {dr\over d\tau}&=&{\Sigma_s \mu\sqrt{\Delta}\over r }u^{(1)}=
\sigma \sqrt{R_s}
\\
\Sigma_s \Lambda_s \mu {d\phi\over d\tau}&=&
{ \Sigma_s  \mu \over r }\left({a\over \sqrt{\Delta}}u^{(0)} +{1+{2Ms^2\over r^3}\over 1-{Ms^2\over r^3}}
u^{(3)}\right)=
\left(1+{3Ms^2\over r\Sigma_s}\right)[
J-(a+s)E]+{a\over \Delta}P_s
\enann
\end{widetext}
where
\beann
P_s&=&\left[r^2+a^2+{as\over r}(r+M)\right] E-\left(a+{Ms\over r}\right)J
\\
R_s&=&P_s^2-\Delta\left[{\mu^2\Sigma_s^2\over r^2}+\left[-(a+s)E+J\right]^2\right]
\,.
\enann
Note that 
\bea
u^{(1)}=\sigma{r\sqrt{R_s} \over \mu \sqrt{\Delta}\Sigma_s }
\,.
\label{specific_radial_momentum}
\ena

Now we introduce the dimensionless variables as
\beann
\tilde E={E\over \mu}\,,~~\tilde J={J\over \mu M}\,,~~\tilde s={s\over  M}
\,,
\enann
\beann
\tilde t={t\over M}\,,~~\tilde r={r\over M}\,,~~a_*={a\over M}\,,~~\tilde \tau={\tau\over M}\,,
\enann
and 
\beann
&&\hskip -2cm
\tilde \Delta=\tilde r^2-2\tilde r+a_*^2\,,
\enann
\beann
&&
\tilde \Sigma_s={\Sigma_s\over M^2}=
\tilde r^2\left(1-{\tilde s^2\over \tilde r^3}\right)\,,
\\
&&
\tilde P_s={P_s\over \mu M^2}
\\
&&
~~~
=\left[\tilde r^2+a_*^2+{a_*\tilde s\over\tilde  r}(\tilde r+1)\right] \tilde E-\left(a_*+{\tilde s\over \tilde  r}\right)\tilde J\,,
\\
&&
\tilde R_s={R_s\over \mu^2 M^4}
\\
&&
~~~=\tilde P_s^2-\tilde \Delta
\left[{\tilde \Sigma_s^2\over \tilde r^2}+\left[-(a_*+\tilde s)\tilde E+\tilde J\right]^2\right]\,.
\enann

The equations of motion are then given by
\beann
\tilde \Sigma_s  
\Lambda_s{d\tilde t\over d\tilde \tau}&=& a_*\left(1+{3\tilde s^2\over \tilde r\tilde \Sigma_s}\right)[
\tilde J-(a_*+\tilde s)\tilde E]+{\tilde r^2+a_*^2\over \tilde \Delta}\tilde P_s
\\
\tilde \Sigma_s \Lambda_s  {d\tilde r\over d\tilde \tau}&=&\pm \sqrt{\tilde R_s}
\\
\tilde \Sigma_s \Lambda_s  {d\phi\over d\tilde \tau}&=& \left(1+{3\tilde s^2\over 
\tilde r\tilde \Sigma_s}\right)[
\tilde J-(a_*+\tilde s)\tilde E]+{a_*\over \tilde \Delta}\tilde P_s
\,.
\enann

\begin{widetext}
\subsection{Constraints on the Orbits}

In what follows, we drop the tilde just for brevity.
In order to find an orbit to the horizon
$r_H:=1+\sqrt{1-a_*^2}$, the radial function $R_s$ must be nonnegative for 
$r\geq r_H$, which 
condition is reduced to be 
\beann
&&
\Big{\{}
\left[ r^3+a_*(a_*+s-b)r+(a_*-b)  s\right]
^2 
-r^2 \Delta
\left(a_*+ s-b\right)^2
\Big{\}}E^2 
\geq   \Delta
 \Sigma_s^2
\,,
\enann
 by introducing an ``impact'' parameter $b:=J/E$.
 There exists a critical value of the impact parameter $b_{\rm cr}$,
beyond which the orbit cannot reach the event horizon.
The particle bounces off at the turning point $dr/d\tau=0$, which  radius is larger than $r_H$. 

The turning point for the critical orbit with $b=b_{\rm cr}$
 is found just at the horizon radius.
 From the condition such that $R_s(r_H)=0$, we find
 \beann
 b_{\rm cr}
 &=&{r_H^3+a_*(a_*+s)r_H+a_*s\over a_*r_H+s}
=
 a_*+s+{r_H^3-s^2\over a_* r_H +s}
 \,.
 \enann
 Hence in order for the orbit to reach the horizon, the condition such that 
 $b\leq b_{\rm cr}$ is required.

There exists  one more important physical condition that  the 4-velocity must be timelike,
which is explicitly written as
\beann
v^\mu v_\mu&=&-(v^{(0)})^2+(v^{(1)})^2+(v^{(3)})^2
={\left[\left(1-X\right)^2\left(
-(u^{(0)})^2+(u^{(1)})^2\right)+\left(1+2 X\right)^2(u^{(3)})^2\right]\over \left[1-X\left(1+3(u^{(3)})^2\right)\right]^2}<0\,,
\enann
where $X={s^2\over r^3}$.
It gives
\beann
\left(1-X\right)^2\left(
-(u^{(0)})^2+(u^{(1)})^2\right)+\left(1+2 X\right)^2(u^{(3)})^2<0
\,.
\enann
\end{widetext}
Since $-(u^{(0)})^2+(u^{(1)})^2+(u^{(3)})^2=-1$, 
this condition is reduced to be
\beann
-(1-X)^2+3X(2+X)(u^{(3)})^2<0
\,.
\enann
From
\beann
u^{(3)}={X^{1/3}\over s^{2/3}(1-X)}[J-(a_*+s)E]
\,,
\enann
 we obtain the  timelike condition of $v^\mu$ as
\bea
{(1-X)^4\over (2+X)X^{5/3}}>{3[J-(a_*+s)E]^2\over s^{4/3}}
\,.
\label{timelike_condition0}
\ena
This condition must be satisfied outside of the event horizon, 
$r\geq r_H$.
Note that the timelike condition is always satisfied for $s=0$.

Since $s^2\leq 1$, $X$ is always smaller than unity outside of the horizon, and
the function on the left hand side in the inequality (\ref{timelike_condition0}) is monotonic 
with respect to $X$, we find the 
above condition is reduced to be
\bea
{(1-X_H)^4\over (2+X_H)X_H^{5/3}}>{3[J-(a_*+s)E]^2\over s^{4/3}}
\,,
\label{timelike_general}
\ena
where $X_H:=s^2/r_H^3$.

By use of the impact parameter $b$, we find
 the above timelike condition for as
\beann
E^2<{s^{4/3}(1-X_H)^4\over 3(b-a_*-s)^2 (2+X_H)X_H^{5/3}}
\,,
\enann
 which gives a constraint on the particle energy $E$.

It is also regarded as a constraint on the impact parameter $b$
for given energy $E$,
i.e.,
\bea
a_*+s-{F(s,r_H) \over E}<b<
a_*+s+{F(s,r_H)\over E}
\label{timelike_condition}
\ena
where 
\beann
F(s,r_H):={s^{2/3}(1-X_H)^2\over \sqrt{3(2+X_H)}X_H^{5/6}}
\,.
\enann

For the critical orbit with $J=J_{\rm cr}$, it becomes 
\bea
E^2<{s^{4/3}(1-X_H)^4\over 3(b_{\rm cr}-a_*-s)^2 (2+X_H)X_H^{5/3}}
\,.
\label{non_extreme_constraint}
\ena

In what follows, we mainly consider the extreme Kerr black hole ($a_*=1, r_H=1$),
especially when we discuss the collisional Penrose process in the next section.
For the extreme black hole, we find $b_{\rm cr}=2$, which does not depend 
on the spin $s$.

If the particle is not critical, by setting $b=2(1+\zeta)$, 
the timelike condition (\ref{timelike_condition}) is rewritten 
as
\bea
&&
-{(1-s)\over 2}-{(1-s^2)^2\over 2E\sqrt{3s^2(2+s^2)}}<\zeta
\nn
&&
~~~~~<-{(1-s)\over 2}+{(1-s^2)^2\over 2E\sqrt{3s^2(2+s^2)}}
\label{timelike_noncritical}
\ena
This gives a constraint on $\zeta$ (or the impact parameter $b=J/E$).

While, for the critical particle with $b_{\rm cr}=2$, from (\ref{non_extreme_constraint})
we have the timelike condition as
\bea
E^2<{(1-s)^2(1+s)^4\over 3s^2 (2+s^2)}
\label{timelike_critical}
\,.
\ena
If the particle plunges from infinity, $E\geq 1$, which gives 
the constraint on the spin $s$ as 
$s_{\rm min} <  s < s_{\rm max}$,
where $s_{\rm min}$ and $s_{\rm max}$ are the solution of the equation
\beann
s^6 + 2 s^5 - 4 s^4- 4s^3 - 7 s^2 + 2 s + 1 = 0
\,,
\enann
with the constraint $s^2\leq 1$.
We find  $s_{\rm min}\approx -0.2709$ and $s_{\rm max}\approx 0.4499$.

\begin{figure}[h]
	\includegraphics[width=6cm]{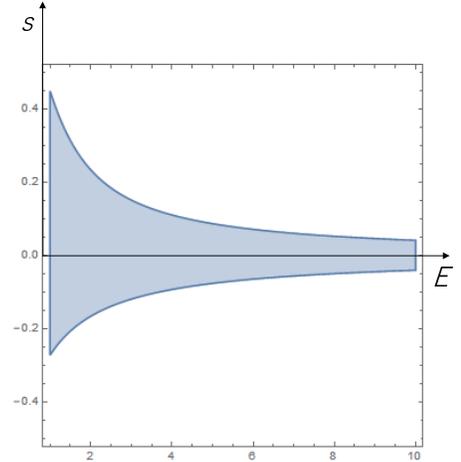}
\caption{The allowed region for the spin $s$ and the energy $E$, with which 
 the particle can reach the event horizon.} 
\label{spinconstraint}
\end{figure}

Eq. (\ref{timelike_critical})  also gives the constraint on a spin $s$ for given 
 particle energy $E$, which is shown in Fig. \ref{spinconstraint}.
This shows the high energy particle cannot reach the horizon if the spin is too large.

When we will discuss a collision in the next chapter, we find that 
the direction of the particle
is important. Since we assume two particles plunge from infinity,
 those particles are ingoing. 
However, if $b>b_{\rm cr}$, a particle falling from infinity will find a turning point, 
and then bounce back to infinity. Such a particle is moving outward.
Hence we consider both directions of the particle motions at collision.

Solving $dr/d\tau=0$ for the angular momentum $J$, we find 
 $J=J_{\pm}(r,E,\mu,s)$, where

\begin{widetext}
\beann
J_{\pm} = {E \{-2 r^4 + r^2 (r^3 - 3r^2 -2) s - r(r + 1) s^2 \} 
	\pm 
	(r-1) (r^3 -s^2) \sqrt{ E^2 r^4 - \mu^2 (r^2 + s) (r^2 - 2r - s)  }
	\over
	r (r^2 + s) (r^2 - 2r - s)
	}\,,
\enann
which gives the bounce point $r$ for a given value of $b=b_\pm:=J_\pm/E$.

\end{widetext}

\begin{figure}[h]
	\includegraphics[width=8.5cm,height=5cm]{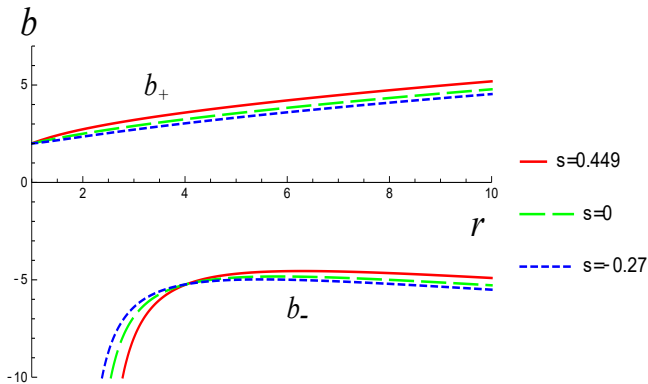}
	\caption{The relation between the turning point $r$  
	 and the impact parameter $b$ for a spinning particle with $E=1$.
	The particle with $b>b_{\rm cr}$ or $b<{\rm max}(b_-)$ 
	 falling from infinity will bounce at the turning point and escape
	 to infinity,
	 while the outward particle with $r<r_{\rm max}$ and 
	 $b<\text{max}(b_-)$ will bounce at the turning point and go back to the horizon, where
	 ${\rm max}(b_-)=-4.97, -4.82, {\rm and~}-4.54$ and
	 $r_{\rm max}=5.48,  5.82, {\rm and~ } 6.30 $ 
	 for $s=-0.27, 0,  {\rm and } 0.449$, respectively.
	 }
	\label{turning_diagram}
\end{figure}

Fig.\ref{turning_diagram} shows 
the turning points for various values of the spin $s$ 
for $E=1$.

We find that ${\rm min}(b_+)=b_{\rm cr}$. 
Then, if the particle is near critical ($b\approx b_{\rm cr}$) but $b>b_{\rm cr}$,
the particle bounces back near the horizon. 

 For the negative value of $b$,  when 
$b<{\rm max}(b_-)$, 
the outgoing particle near the horizon will bounce back to the horizon, while the particle coming from infinity will bounce back to infinity.
We find 
${\rm max}(b_-) \approx -4.97,  -4.82, {\rm and~} -4.54$ 
for $s=-0.27, 0, {\rm and~} 0.449$,
respectively.

\begin{figure}[h]
	\includegraphics[width=6cm]{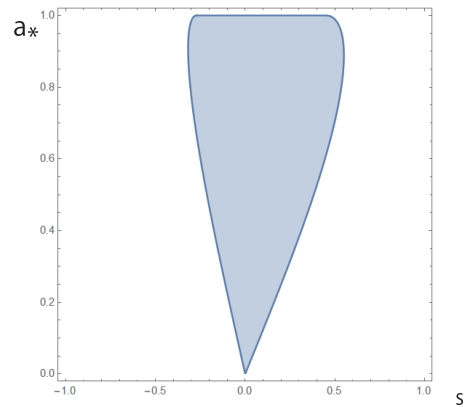}
\caption{The parameter region $(s,a_*)$ for the existence of the timelike critical orbit with $E\leq 1$ 
until the event horizon.} 
\label{timelike}
\end{figure}

For nonextreme black hole, 
 from Eq. (\ref{non_extreme_constraint}), 
 the timelike condition  for the critical orbit 
with $E\geq 1$ gives the necessary conditions on the parameters of $(s,a_*)$,
which is shown in Fig. \ref{timelike}.
For $a_*=0.9$, $E\geq 1$ gives $-0.3179<  s <0.5497$,
which range is a little larger than the extreme case.
While for $a_*=0$ (Schwarzschild black hole), no region exists
because there is no critical orbit.

\subsection{Orbit of a massless particle on the equatorial plane}
Since we also discuss the scattering of massless particle later, we 
shall describe its orbit on the equatorial plane in the Kerr geometry.
A  massless particle particle is not spinning ($s=0$).
Hence, the conserved  energy and the $z$-component of the angular momentum 
of the massless particle are defined by 
\beann
E=-p^\mu \xi^{(t)}_{\mu}\,,~{\rm and}~~J=p^\mu\xi^{(\phi)}_\mu
\,.
\enann
Then we find
\beann
p^{(0)}=
{\left[
(r^2+a^2)E-a)J\right]\over  r\sqrt{\Delta}}
\,,~{\rm and}~~
p^{(3)}=
{\left[
J-aE\right]\over  r}
\,.
\enann
This gives 
\beann
p^{(1)}&=&\sigma\sqrt{(p^{(0)})^2-(p^{(3)})^2}
\\
&=&{\sigma\over r\sqrt{\Delta}}\sqrt{\left[
((r^2+a^2)E-aJ)^2-(J-aE)^2\Delta\right]}
\,.
\enann
When we discuss the orbit we have to look at the 4-velocity $v^\mu={dz^\mu\over d\lambda}$, where $\lambda$ is an affine parameter.
The 4-momentum $p^\mu$ and the 4-velocity $v^\mu$ are proportional.
By choosing the affine parameter $\lambda$ appropriately, 
we can set 
\beann
p^\mu=E v^\mu\,.
\enann
As a result, we find 
\beann
\left({dr\over d\lambda}\right)^2&=&{\Delta\over r^2}\left(v^{(1)}\right)^2
=
{\Delta\over r^2}{\left(p^{(1)}\right)^2\over E^2}
\\
&=&
{1\over r^4 E^2}\left[
((r^2+a^2)E-aJ)^2-(J-aE)^2\Delta\right]
\,.
\enann

Using the ``impact'' parameter $b=J/E$, we find the critical value 
\beann
b_{\rm cr}={r_H^2+a^2\over a}={2Mr_H\over a}
\,,
\enann
beyond which the photon orbit bounces before the horizon.
For the extreme black hole, we find the same critical value $b_{\rm cr}=2$ as that for the massive particle.

\section{Collision of Spinning Particles}
\label{collision}
Now we discuss the collision of two particles moving in extreme Kerr geometry
($a_*=1$), in which we expect the maximal energy extraction.
Two particles 1 and 2, whose 4-momenta are $p_1^\mu$ and
$p_2^\mu$, are moving to a rotating black hole
and collide just before 
the horizon.  After the collision, the particles 3 with the 4-momentum $p_3^\mu$ 
is going away to infinity, while the particle 4 with the 4-momentum $p_4^\mu$ falls into the black hole. 

We assume that 
the sum of two momenta and spins, if any,  are conserved at the collision, i.e.,
\beann
p_1^\mu+p_2^\mu&=&p_3^\mu+p_4^\mu
\\
S_1^{\mu\nu}+S_2^{\mu\nu}&=&S_3^{\mu\nu}+S_4^{\mu\nu}
\,.
\enann
From those conservations with the Killing vectors, 
we find the conservations of the energy and total angular momentum, 
\beann
E_1+E_2&=&E_3+E_4
\\
J_1+J_2&=&J_3+J_4
\,.
\enann
We also obtain that the sum of the spins and the radial components of  
4-momenta are conserved 
at the collision; 
\beann
\mu_1 s_1+\mu_2 s_2&=&\mu_3 s_3+\mu_4 s_4
\\
p_1^{(1)}+p_2^{(1)}&=&p_3^{(1)}+p_4^{(1)}
\,.
\enann

In what follows, we discuss two cases: {\bf [A]} collision of two massive particles
({\bf MMM}), and 
{\bf [B]}  collision of massless and massive particles 
; the Compton scattering ({\bf PMP}) and inverse Compton scattering ({\bf MPM}),
where we use the symbols of {\bf MMM}, {\bf PMP}, and {\bf MPM} following \cite{Leiderschneider:2015kwa}.
{\bf P} and {\bf M} describe a massless particle (a photon) and a massive particle, respectively.
The first and the second letters denote colliding particles, while 
the third letter shows an escaped particle.

For the case {\bf [A]}{\bf MMM}, 
we assume all masses of the particles are the same, i.e.,
$\mu_1=\mu_2=\mu_3=\mu_4=\mu$.
Hence the conservation equations hold for the dimensionless specific variables:
\bea
\tilde E_1+\tilde E_2&=&\tilde E_3+\tilde E_4
\label{collision_condition1}
\\
\tilde J_1+\tilde J_2&=&\tilde J_3+\tilde J_4
\label{collision_condition2}
\\
\tilde s_1+\tilde s_2&=&\tilde s_3+\tilde s_4
\label{collision_condition3}
\\
u_1^{(1)}+u_2^{(1)}&=&u_3^{(1)}+u_4^{(1)}
\,.
\label{collision_condition4}
\ena

For the case {\bf [B]}{\bf PMP}, we assume that 
the particles 1 and 3 are massless and nonspinning, corresponding to a photon, 
while the particles 2 and 4  have the same mass, i.e., $\mu_2=\mu_4=\mu$.
We then have 
\bea
\tilde s_2&=&\tilde s_4
\label{Compton_condition3}
\\
p_1^{(1)}+p_2^{(1)}&=&p_3^{(1)}+p_4^{(1)}
\,,
\label{Compton_condition4}
\ena
in addition to two conservation equations (\ref{collision_condition1}) and (\ref{collision_condition2}).
In the case of  {\bf [B]}{\bf MPM}, the particles 2 and 4 are massless and nonspinning, 
while the particles 1 and 3  are massive with the same mass, i.e., $\mu_1=\mu_3=\mu$,
and Eq. (\ref{Compton_condition3}) is replaced by 
\bea
\tilde s_1&=&\tilde s_3
\,.
\label{inverse_Compton_condition3}
\ena

As we showed, there exists a critical orbit, 
which satisfies $J=J_{\rm cr}=2 E$ in the extreme Kerr spacetime.
This orbit will reach to the event horizon, and then bounce there.
If $J<J_{\rm cr}$, the orbit gets into a black hole.
While when $J>J_{\rm cr}$,
the orbit  bounces back before the horizon.

We assume that 
 the particles 1 and 2 starting from infinity are falling toward a black hole, and collide near the event horizon, i.e., 
 the collision point $r_c$ is very close to the horizon ($r_H=1$), i.e., 
$r_c=1/(1-\epsilon)$
($0<\epsilon \ll 1$).
Hence the leading order of the radial component of the 
4-momentum $p^{(1)}$ is
\beann
p^{(1)}\approx \sigma {|2E-J|\over \epsilon(1-s)}+\cdots
\,.
\enann

The momentum conservation equation ($p_1^{(1)}+p_2^{(1)}=p_3^{(1)}+p_4^{(1)}$) yields

\begin{widetext}
\bea
&&
\sigma_1 {|2E_1-J_1|\over 1-s_1}
+\sigma_2 {|2E_2-J_2|\over 1-s_2}
=\sigma_3{|2E_3-J_3|\over 1-s_3}
+\sigma_4{|2E_4-J_4|\over 1-s_4}
+O(\epsilon)
\label{leading_order}
\ena

\end{widetext}
In what follows, we consider just the
 case such that
 the particle 1 is critical ($J_1=2E_1$). 

To classify the case, we consider 
 two situations for the particle orbits:
One is near-critical ($J=2 E+O(\epsilon)$), and the other is noncritical
($J=2 E+O(\epsilon^0)$).  
Since we consider the collision near the horizon,
noncritical orbit must have a smaller angular momentum $J<2E$.

From Eq. (\ref{leading_order}), we find the following four cases:
\\
(1) Both particle 2 and particle 3 are near-critical. In this case there is no constraint on $\sigma_2, \sigma_3$ and $\sigma_4$. 
\\
(2) The  particle 2 is near-critical but the particle 3 is noncritical ($J_3<2E_3$). In this case, 
using the conservation equations (\ref{collision_condition1}) and (\ref{collision_condition2}),
we find 
\beann
\left[{\sigma_3\over 1-s_3}+{\sigma_4
\over 1-s_4}\right]
(J_3-2 E_3)=
O(\epsilon)
\,.
\enann
We find $\sigma_4=-\sigma_3$ and $s_4=s_3=s$. For the case {\bf [B]},
 since $s_3=0$ or $s_4=0$, the massive particles are also nonspinning.
\\
(3) The  particle 3 is near-critical but the particle 2 is noncritical
($J_2<2E_2$). In this case, we find 
\beann
\left[{\sigma_4\over 1-s_4}-{\sigma_2 \over 1-s_2}\right]
(J_2-2 E_2)=
O(\epsilon)
\,.
\enann
We find $\sigma_4=\sigma_2$ and $s_4=s_2$. Hence 
we have to impose $s_3=s_1$. 
\\
(4) Both particle 2 and particle 3 are noncritical
($J_2<2E_2$ and $J_3<2E_3$). 
In this case there is no constraint on $\sigma_2, \sigma_3$ and $\sigma_4$.

Here  we shall analyze only the case (3).
It is because it gives a good efficiency as we will show below.
We will  not discuss the other three cases (1), (2) and (4) 
in this paper.
It is because it does not
seem to get a good efficiency for the cases (1) and (2).
For the case (4), the super-Penrose process could be possible, but it is not possible to
 analyze it by our present method.

Since we consider the collision of the particle 1 and the particle 2, 
the noncritical particle 2 with $J_2<2E_2$ must be ingoing ($\sigma_2=-1$).
So we assume that $\sigma_4=\sigma_2=-1$.
While 
the critical particle 1 can be either ingoing ($\sigma_1=-1$) or outgoing after 
  a bounce near the horizon ($\sigma_1=1$). 
The latter case is not exactly correct. 
In order for the particle 1 to bounce, 
it must be supercritical such that  $J_1=2E_1+\delta$ with
 $\delta>0$. We then take a limit of $\delta\rightarrow 0$,
 which gives the ``critical orbit'' with a bounce.
 Since we also have a
small parameter $\epsilon$, we have to take a limit of $\delta\rightarrow 0$ first, which 
implies  $\delta\ll \epsilon$.

The above setting gives 
\bea
J_1&=&2 E_1
\\
J_3&=&2 E_3 (1+\alpha_3\epsilon+\beta_3 \epsilon^2+\cdots)
\,,
\label{case3_1}
\ena
where $\alpha_3$  and $\beta_3$ are parameters of $O(\epsilon^0)$.

As for the particle 2, 
 we assume 
\bea
J_2&=&2 E_2 (1+\zeta)
\,,
\label{case3_2}
\ena
where $\zeta<0$ with  $\zeta=O(\epsilon^0)$.

From the conservation laws, 
we find 
\bea
E_4=E_1+E_2-E_3\,,~~~J_4=J_1+J_2-J_3
\,,
\label{conservation}
\ena
giving 
\beann
J_4=2 E_4\left(1+{E_2\over E_4}\zeta
+\cdots\right)
\,.
\enann

Now we evaluate  $E_2$ and $E_3$ for the cases {\bf [A]} and {\bf [B]} separately.

\begin{widetext}
\subsection{Case {\bf [A] MMM} {\rm (Collision of two massive particles)}}

For the massive particle, the radial component of the 
specific 4-momentum is written as
\bea
&&u^{(1)}=\sigma {r\sqrt{R_s}\over \Sigma_s\sqrt{\Delta}}
\nonumber 
\\
&&
={\sigma\sqrt{r^2\left[(r^3+(1+s)r+s)E-(r+s)J\right]^2-(r-1)^2\left[(r^3-s^2)^2+r^4(J-(1+s)E)^2\right]}
\over (r-1)(r^3-s^2)}
\label{radial_momentum}
\,.
\ena
Plugging the conditions (\ref{case3_1}) and (\ref{case3_2})
into Eq. (\ref{radial_momentum}),
and using the conservation equations (\ref{conservation})
we find
\bea
u_1^{(1)}&=&\sigma_1\Big[{f(s_1,E_1,0)\over (1 - s_1^2)}
-\epsilon{E_1^2 h(s_1)\over (1 - s_1^2)^2 
f(s_1,E_1,0)}+O(\epsilon^2)
\Big]
\label{momentum1}
\\
u_2^{(1)}&=&\epsilon^{-1}{2 E_2 (1+s_2) \zeta\over 1-s_2^2}
-{E_2(2+s_2)(1-s_2+2\zeta)\over (1-s_2)^2(1+s_2)}
\nn
&&
-\epsilon {(1-s_2)^4(1+s_2)^2+E_2^2\left(1-s_2+2\zeta\right)\left[(1-s_2)^3-2(1+2s_2)(1+4s_2+s_2^2)\zeta
\right]\over 4 (1 - s_2)^3 (1 + s_2)^2 E_2\zeta}+O(\epsilon^2)
\label{momentum2}
\ena
\bea
u_3^{(1)}&=&\sigma_3\Big{\{}{f(s_1,E_3,\alpha_3)\over (1 - s_1^2)}
-\Big{[}{\epsilon E_3^2\over (1-s_1^2)^2f(s_1,E_3,\alpha_3)}
\times
 \Big(h(s_1)
 - 2 (1 + s_1)^2(2 + s_1) g_2(s_1,\alpha_3)
\nn
&&
 +2\beta_3 (1+s_1)  (1 - s_1^2) g_1(s_1,\alpha_3) \Big)
\Big{]}+O(\epsilon^2)
\Big{\}}
\label{momentum3}
\\
u_4^{(1)}&=&\epsilon^{-1}{2 E_2 (1+s_2) \zeta\over 1-s_2^2}
-{[E_1 (1-s_2)(2+s_2) -E_3 (1- s_2)  g_1(s_2,\alpha_3)+ 
   E_2 (2 + s_2) (1 - s_2 + 2 \zeta)] \over (1 - s_2)^2  (1 + s_2) }
\nn
   &&
-{\epsilon\over 4 (1 - s_2)^3 (1 + s_2)^2 E_2   \zeta }\Big{[}
   (1 - s_2)^4 [(E_1-E_3)^2 + (1 + s_2)^2]
\nn
 &&
  - 2 E_2  (1 - s_2)\{4(1+s_2)E_3\zeta[
 \alpha_3 (2 +s_2)- \beta_3 (1 - s_2^2)  ] 
  + (E_3-E_1 )
    [(1-s_2)^3-2s_2(2+s_2)^2\zeta]\}
    \nn
 &&
 + 
 E_2^2(1-s_2+2\zeta)[
 (1-s_2)^3-2 (1+2s_2)(1+4s_2+s_2^2)\zeta] 
    \Big{]}
  +O(\epsilon^2)
 \label{momentum4}
 \,,
\ena
where
\beann
f(s,E,\alpha)&:=&\sqrt{E^2 [3 - 2 \alpha (1+s)][1 + 2 s - 2 \alpha (1+s)] - (1 -  s^2)^2}
\,,
\\
g_1(s,\alpha)&:=&2+s-2\alpha(1+s)
\,,
\\
g_2(s,\alpha)&:=&\alpha (2 + s-2  \alpha)
\,,
\\
h(s)&:=&1 + 7 s + 9 s^2 + 11 s^3 - s^4
\enann

Since $u_1^{(1)}+u_2^{(1)}=u_3^{(1)}+u_4^{(1)}$, 
we find the leading order of $\epsilon^{-1}$ is trivial.
From the next leading order of $\epsilon^0$, we find
\beann
&&
\sigma_3{f(s_1,E_3,\alpha_3)\over 1-s_1^2}
=\sigma_1{ f(s_1,E_1,0)\over 1-s_1^2}
+{\left[E_1 (2+s_2) -E_3 g_1(s_2,\alpha_3)
\right]\over 1-s_2^2}
\,,
\enann
which is reduced to
\bea
{\cal A}E_3^2-2{\cal B}E_3+{\cal C}=0
\,,
\label{eq_E3}
\ena
where
\bea
{\cal A}&=&
-[3-2\alpha_3(1+s_1)][1+2s_1-2\alpha_3(1+s_1)]
+{(1-s_1^2)^2\over (1-s_2^2)^2}g_1^2(s_2,\alpha_3)
\label{calA}
\\
{\cal B}&=&g_1(s_2,\alpha_3){(1-s_1^2)\over (1-s_2^2)}\left[
(2+s_2){(1-s_1^2)\over (1-s_2^2)}
E_1+\sigma_1f(s_1,E_1,0)
\right]
\label{calB}
\\
{\cal C}&=&E_1\left[
\left({3(1+2s_1)(1-s_2^2)^2+(1-s_1^2)^2(2+s_2)^2\over (1-s_2^2)^2}\right) E_1
+2\sigma_1{(1-s_1^2)(2+s_2)\over (1-s_2^2)} f(s_1,E_1,0)
\right]
\label{calC}
\,,
\ena
with the condition such that 
$E_3\leq E_{3, {\rm cr}}$ for $\sigma_3=1$, or  
$E_3\geq E_{3, {\rm cr}}$ for $\sigma_3=-1$,
where
\beann
E_{3, {\rm cr}}:={1\over g_1(s_2,\alpha_3)}\left[
(2+s_2)E_1+\sigma_1{(1-s_2^2)\over (1-s_1^2)}f(s_1,E_1,0)
\right]
\,.
\enann

Here we focus just into the case of $\sigma_3=-1$.
We should stress that for the outgoing particle 3 after collision ($\sigma_3=1$),
the energy $E_3$ has the upper bound $E_{3, {\rm cr}}$, which magnitude is the order of 
$E_1$. Hence we may not expect large efficiency.
We will present the concrete analysis for the case of $\sigma_3=1$ in Appendix
\ref{appendix}, in which we find the efficiency is not so high.

Since the particle 3 is ingoing after the collision, the orbit must be supercritical,
i.e., $J_3>2 E_3$, which means either $\alpha_3 >0$ or $\alpha_3=0$ with $\beta_3>0$.
Once we give $\alpha_3$, the  value of $E_3$ is fixed 
in terms of $s_1$, $s_2$ and $E_1$ by
\bea
E_3&=&E_{3,+}:={{\cal B}+\sqrt{{\cal B}^2-{\cal A}{\cal C}} \over {\cal A}}
\label{energy_E3}
\,,
\ena
where we have chosen the larger root because it gives the larger extracted energy
as it turns out that  ${\cal A}$ is always positive.

The next leading order terms give
\bea
{\cal P}E_2
   =      (1 - s_2)^3 (E_1-E_3)^2
   \,,
\label{energy_E2}
\ena
where
\bea
{\cal P}:&=& 2 (E_3-E_1 )(1-s_2)^3+ 4\zeta\Big[
 {(1 - s_2^2)^2 \over (1-s_1^2)^2}   {\cal Q}
  +  2(1+s_2)E_3[
 \alpha_3 (2 +s_2)- \beta_3 (1 - s_2^2)  ] 
  -s_2(2+s_2)^2 (E_3-E_1 )
\Big]
\,.~~~
\label{calP}
\ena
with
\beann
{\cal Q}&:=&
\sigma_1{E_1^2 h(s_1)\over  
f(s_1,E_1,0)}
-\sigma_3\Big{[}{ E_3^2\over f(s_1,E_3,\alpha_3)}
\times
 \Big(h(s_1)
 - 2 (1 + s_1)^2(2 + s_1)g_2(s_1,\alpha_3)
 +2\beta_3 (1+s_1)  (1 - s_1^2)g_1(s_1,\alpha_3)\Big)
\Big{]}
\label{calQ}
\enann

Since this fixes the value of  $E_2$, we obtain
the efficiency  by 
\beann
\eta={E_3\over E_1+E_2}
\,,
\enann
when  $\alpha_3, \beta_3$ and $\zeta$ are given.

\subsection{Case {\bf [B]}}
\subsubsection{{\bf [B] PMP} {\rm (Compton scattering)}}
For the massless particle, we normalize the 4-momentum, the energy and the angular momentum  by the mass  $\mu$ of the massive particle.
The radial component of the normalized 
 4-momentum is written as
\bea
p^{(1)}
={\sigma\sqrt{r\left[(r+1)E-J\right]\left[(r^2-r+2)E+(r-2)J\right]}
\over r(r-1)}
\label{massless_radial_momentum}
\,,
\ena
where $E$ and $J$ are normalized by $\mu$ and $\mu M$ just as those of the massive particle.

For the momenta of the massive particles 2 and 4, Eqs.  (\ref{momentum2}) and (\ref{momentum4}) do not change, 
while for the massless particles 1 and 3, we find
\bea
p_1^{(1)}&=&\sigma_1\Big[\sqrt{3}E_1
-\epsilon{E_1 \over  
\sqrt{3}}+O(\epsilon^2)
\Big]
\label{photon1}
\\
p_3^{(1)}&=&\sigma_3\Big{\{}E_3\sqrt{(3-\alpha_3)(1-2\alpha_3)}
-\epsilon E_3\Big{[}{  [1
 - 4(2\alpha_3 
-\beta_3)  (1- \alpha_3) ] \over \sqrt{(3-2\alpha_3)(1-2\alpha_3)}}
\Big{]}+O(\epsilon^2)
\Big{\}}
 \label{photon3}
 \,.
\ena

From the conservation of the radial components of the 4-momenta, 
we find 
\bea
&&
E_3={\cal S}E_1
\,,
\label{Compton_energy_E3}
\ena
where the magnification factor ${\cal S}$ is given by 
\beann
{\cal S}:={\sigma_1\sqrt{3}(1-s_2^2)+2+s_2
\over\sigma_3\sqrt{(3-2\alpha_3)(1-2\alpha_3)}(1-s_2^2)
+2+s_2-2\alpha_3(1+s_2)
}
\label{Compton_magnification}
\enann
and 
\bea
{\cal P}E_2
   =      (1 - s_2)^3 (E_1-E_3)^2
   \,,
\label{Compton_energy_E2}
\ena
where ${\cal P}$ is given by Eq. (\ref{calP}) with $s_1=0$ but replacing ${\cal Q}$ by ${\cal T}$,
which is defined by
\beann
{\cal T}&:=&
\sigma_1{E_1 \over  
\sqrt{3}}
-\sigma_3 E_3\Big{[}{  1
 - 4(2\alpha_3 
-\beta_3)  (1- \alpha_3) \over \sqrt{(3-\alpha_3)(1-2\alpha_3)}}
\Big{]}
\,.
\enann

\subsubsection{Case {\bf [B] MPM} {\rm (Inverse Compton scattering)}}
For the momenta of the massive particles 1 and 3, Eqs.  (\ref{momentum1}) and (\ref{momentum3}) 
do not change, while for the massless particles 2 and 4, we find
\bea
p_2^{(1)}&=&2\epsilon^{-1}E_2\zeta -2E_2(1+2\zeta)
-\epsilon{E_2 (1-4\zeta^2)\over  4\zeta}+O(\epsilon^2)
\label{photon2}
\\
p_4^{(1)}&=&2\epsilon^{-1}E_2\zeta-2\left[E_4+2E_2\zeta+E_3\alpha_3\right]
-\epsilon{E_4^2-8E_2E_3(2\alpha_3-\beta_3)\zeta-4E_2^2\zeta^2\over 4 E_2\zeta}
+O(\epsilon^2)
 \label{photon4}
 \,.
\ena
\end{widetext}
where $E_4=E_1+E_2-E_3$

From the conservation of the radial components of the 4-momenta, 
we find 
\bea
E_3&=&
\left. 
{{\cal B}+\sqrt{{\cal B}^2-{\cal A}{\cal C}} \over {\cal A}}
\right|_{s_2=0}
\label{Inverse_Compton_energy_E3}
\,,
\ena
and 
\bea
E_2&=&
   \left.{(E_1-E_3)^2 \over {\cal P}} \right|_{s_2=0}
   \,,
\label{Inverse_Compton_energy_E2}
\ena
where ${\cal A}, {\cal B}, {\cal C}$ and ${\cal P}$ are given by Eqs.
 (\ref{calA}),  (\ref{calB}),  (\ref{calC}) and (\ref{calP}),
 which should be evaluated with $s_2=0$.
As a result, $E_2$ and $E_3$ coincide with
those found at the  collision of 
a spinning massive particle and a nonspinning massive particle.

\section{The maximal efficiency}
\label{max_efficiency}
\subsection{Efficiency of Collision of Massive Particles}
Now we discuss  the necessary condition to find  the maximal efficiency.
As we showed, giving the particle 1 energy  ($E_1$)
and two particle spins ($s_1$ and $s_2$), we find the energies of the  particle 3
and particle 2  in terms of the orbit parameters of the particles 2 and 3 ($\alpha_3$, $\beta_3$ and $\zeta$).
In order to get the large efficiency, we must find large extraction energy, i.e., the energy of the particle 3 ($E_3$)
for given values of  $E_1$ and $E_2$ of  the ingoing particles.
Although $E_1$ is arbitrary,  
the energy of the particle 2 ($E_2$)  is fixed 
in our approach.  Hence we also have to find 
 the possible minimum value of $E_2$.  Since we consider two particles are plunging from infinity, we have the constraints of 
$E_1\geq 1$ and $E_2\geq 1$. 

We then assume that $E_1=1$ and $\sigma_1=1$, 
and find the maximal value of $E_3$ as well as the minimum value of $E_2$.
Note that  we do not find a good efficiency for $\sigma_1=-1$, although 
the off-plane orbits may give a little better efficiency\cite{Leiderschneider:2015kwa}.

First we analyze $E_3$, which is determined by 
Eq. (\ref{energy_E3}) for given value of  $\alpha_3$.
Since the orbit of the particle 3 is near critical, 
we have two constraints:
$E_3\geq E_{3, {\rm cr}}$ for $\sigma_3=-1$
and 
the timelike condition (\ref{timelike_critical}).

\begin{figure}[h]
	\includegraphics[width=6cm]{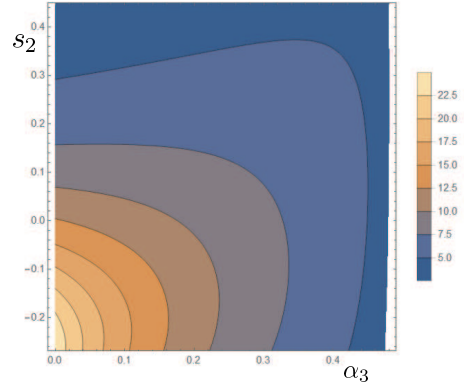}
\caption{The contour map of $E_3$ in terms of $\alpha_3$ and $s_2$ with $s_1=0$.
$E_3$ changes smoothly with respect to two parameters $\alpha_3$ and $s_2$,
and $\alpha_3\rightarrow 0$ and small $s_2$ give larger value of $E_3$.} 
\label{a3s2}
\end{figure}

In order to find the large value of $E_3$, from the timelike condition
we find that 
the spin magnitude $s_3(=s_1)$ must be small
(see Fig. \ref{spinconstraint}).
Hence we first set $s_1=0$.
We then show the contour map of 
$E_3$ in terms of $\alpha_3$ and $s_2$
in Fig. \ref{a3s2}.	
We find $\alpha_3 \approx 0$ gives the largest efficiency.
Hence next  we set $\alpha_3=0+$, and analyze the maximal efficiency.
Here $0+$ means that
 we assume $\alpha_3>0$ but take a limit of $\alpha_3\rightarrow 0$ after 
taking the limit of $\epsilon\rightarrow 0$.
This is justified because $E_2$ and $E_3$ change smoothly when we take the limit of 
$\alpha_3\rightarrow 0$.

\begin{figure}[h]
	\includegraphics[width=6cm]{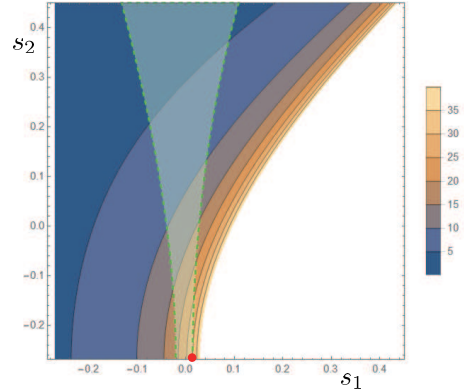}
\caption{The contour map of $E_3$ in terms of $s_1$ and $s_2$.
The timelike condition for the particle 3 orbit is satisfied in the light green shaded region.
As a result, the maximal value of $E_3=E_{3,{\rm max}}\approx 30.02$  is obtained when $s_2=s_{\rm min}\approx-0.2709$ and $s_1\approx 0.01379$ (the red point in the figure).} 
\label{maxE3}
\end{figure}

Assuming $\alpha_3=0+$,
 we look for the maximal value of $E_3$ for given $s_1$ and $s_2$.
In Fig.\ref{maxE3}, we show the contour map of $E_3$ in terms of 
$s_1$ and $s_2$.
The red point, which is $(s_1, s_2)\approx (0.01379, s_{\rm min})$,
 gives the maximal value of $E_3$.

Since $E_2\geq 1$ when we plunge the particle 2 from infinity, 
if $E_2=1$ is possible, we find that the maximal value of $E_3$ gives the maximal efficiency.
However $E_2$ is fixed in our approach. 
So we have to check whether  $E_2=1$ is possible or not 
and then provide which conditions are required if possible. 

\begin{figure}[h]
	\includegraphics[width=4cm]{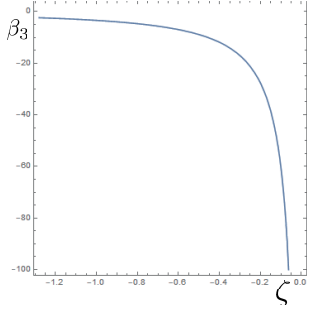}
\caption{The relation between $\zeta$ and $\beta_3$ for $E_2=1$.
The other parameters are chosen to give the maximal value of $E_3$.
The timelike condition for the particle 2 orbit
gives  the constraint of $\zeta_{\rm min}<\zeta<0$
with $\zeta_{\rm min}\approx -1.271$.} 
\label{E2}
\end{figure}

\begin{figure}[h]
	\includegraphics[width=6cm]{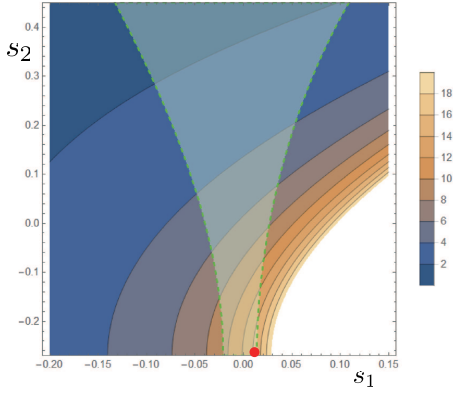}
\caption{The contour map of the maximal efficiency for given $s_1$ and $s_2$.
The green shaded region is the constraint from the timelike condition of the particle 3.
The red point, $(s_1, s_2)\approx (0.01379, s_{\rm min})$, gives the maximal efficiency $\eta_{\rm max}\approx 15.01$.} 
\label{efficiency}
\end{figure}

\begin{figure}[h]
\vskip .5cm
\includegraphics[width=5cm]{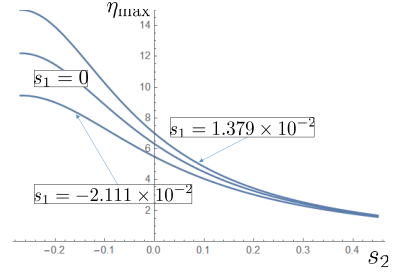}
\caption{The  efficiency in terms of $s_2$ for  fixed values of 
$s_1=-2.111\times 10^{-2}, 0$ and $1.379\times 10^{-2}$.} 
\label{eff_max}
\end{figure}

The condition for $E_2=1$ in Eq. (\ref{energy_E2}) gives the relation between $\zeta$ and $\beta_3$, which is a linear equation of $\beta_3$. Hence we always find a real solution of $\beta_3$.
While the timelike condition of the particle 2 gives the constraint on $\zeta$,
which is Eq. (\ref{timelike_noncritical}) with $E=1$, i.e.,
\beann
\zeta_{\rm min}<\zeta
<0
\,,
\label{timelike_E2}
\enann
where 
\beann
\zeta_{\rm min}:= -{(1-s_2)\over 2}\left[1+{(1-s_2)(1+s_2)^2\over \sqrt{3s_2^2(2+s_2^2)}}\right]
\enann
since the upper bound in  Eq. (\ref{timelike_noncritical})  is always positive for the range of 
$s_{\rm min}<s_2<s_{\rm max}$.
For the parameters giving the maximal value of $E_3$, 
we find the relation 
between $\zeta$ and $\beta_3$, which is shown in Fig. \ref{E2}. 
From the timelike condition for the particle 2 orbit,
we have the constraint of $\zeta_{\rm min}<\zeta<0$
where $\zeta_{\rm min}\approx -1.271$.

Since there exists a possible range of parameters with $E_2=1$, 
we find the maximal efficiency is
given by $\eta_{\rm max}=E_{3,{\rm max}}/2\approx 15.01$.

Hence we find the maximal efficiency $\eta_{\rm max}=E_3/2$ for given 
$s_1$ and $s_2$,
which is shown in Fig. \ref{efficiency}.
We also show the efficiency in terms of $s_2$ for fixed values of $s_1=-2.111\times 10^{-2}, 0$ and $1.379\times 10^{-2}$ in Fig. \ref{eff_max}.
The efficiency gets larger as $s_2$ approaches the minimum value $s_{\rm min}$.
It shows that the effect of spin is very important.
Note that we obtain the maximal efficiency $\eta_{\rm max}\approx 6.328$
for nonspinning case, which is 
consistent with \cite{Leiderschneider:2015kwa}.

\subsection{Efficiency of Compton scattering}
We find the efficiency $\eta$ by
\beann
\eta&=&{E_3\over E_1+E_2}
=
{{\cal S}\over 1+{({\cal S}-1)^2(1-s_2)^3\over {\cal P}/E_1}}
\enann
where 
\beann
{\cal P}/E_1&=&2({\cal S}-1)(1-s_2)^3
\\
&+&
4\zeta\Big{[}
(1-s_2^2)^2{\cal T}/E_1+2(1+s_2){\cal S}
[\alpha_3(2+s_2)
\\
&-&
\beta_3(1-s_2^2)]-s_2(2+s_2)^2({\cal S}-1)
\Big{]}
\enann
with 
\beann
{\cal T}/E_1={\sigma_1\over \sqrt{3}}-\sigma_3{\cal S}
\Big{[}{1 - 4(2\alpha_3 -\beta_3)  (1- \alpha_3) \over \sqrt{(3-\alpha_3)(1-2\alpha_3)}}
\Big{]}
\,.
\enann
Although the extracted photon energy depends on  the input photon energy $E_1$, the efficiency does not depend on $E_1$ and $E_2$.  It is determined by 
the orbital parameters $\alpha_3$, $\beta_3$ and $\zeta$ as well as the spin $s_2$.

\begin{figure}[h]
\includegraphics[width=6cm]{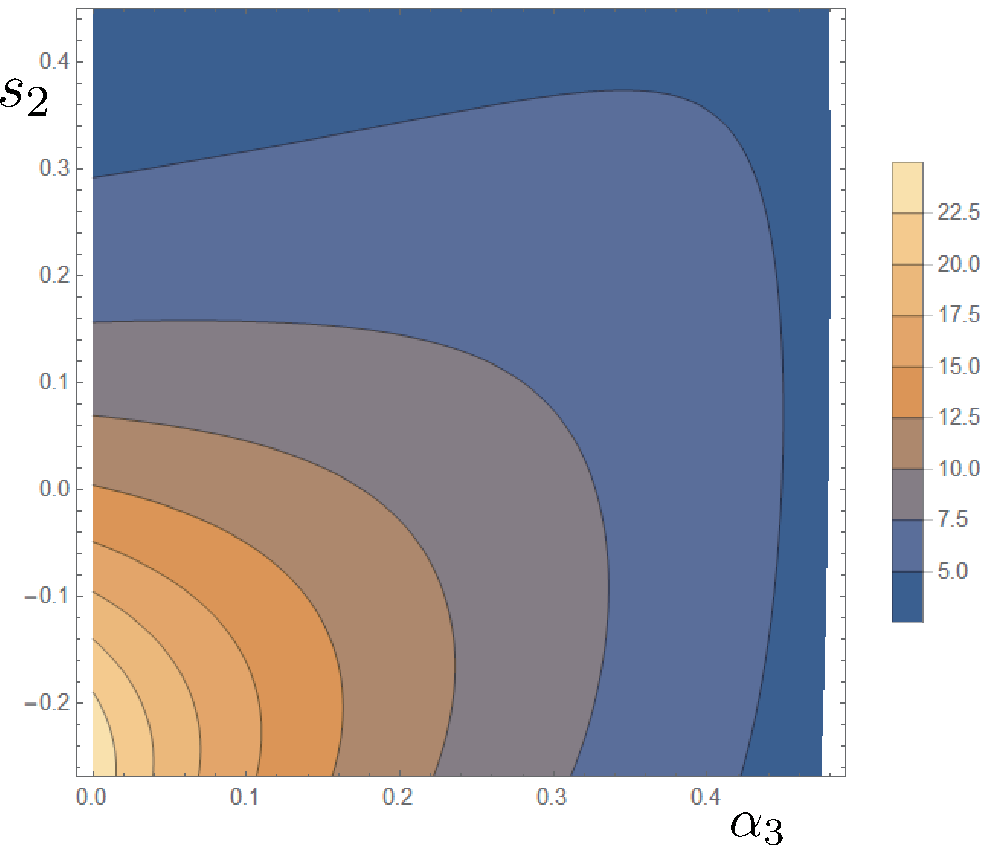}
\caption{The contour map of ${\cal S}$ in terms of $\alpha_3$ and $s_2$.
${\cal S}$ changes smoothly with respect to two parameters $\alpha_3$ and $s_2$,
and $\alpha_3\rightarrow 0$ and small $s_2$ give larger value of ${\cal S}$.} 
\label{Compton_E3}
\end{figure}

We first look for when we find the largest value of $E_3$,
or the magnification factor ${\cal S}$, 
which is determined by $\alpha_3$.
In Fig. \ref{Compton_E3}, we show the magnification factor ${\cal S}$ in terms of
$\alpha_3$ and $s_2$. Just as the case {\bf [A]},
$\alpha_3\rightarrow 0$ and small $s_2$ give larger value of ${\cal S}$.
The maximal value is ${\cal S}_{\rm max}\approx 26.85$ at  $\alpha_3=0+$
and $s_2=s_{\rm min}\approx -0.2709$.

\begin{figure}[h]
	\includegraphics[width=6cm]{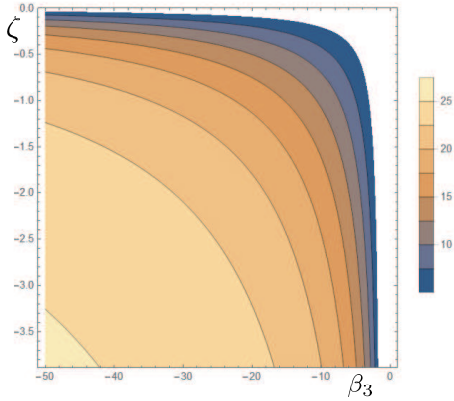}
\caption{The contour map of the efficiency in terms of $\beta_3$ and $\zeta$.
Fixing  $\zeta$ with $0>\zeta>\zeta_{\rm min}(\approx -3.890)$,
 in the limit of  $\beta_3\rightarrow \infty$, 
we find  the maximal efficiency of $\eta_{\rm max}\approx 26.85$.} 
\label{Compton_eff}
\end{figure}

Since  the maximal value of ${\cal S}$ is obtained when  $\alpha_3\rightarrow 0$ and $s_2=s_{\rm min}$,
setting $\alpha_3=0+$ and $s_2=s_{\rm min}$, we show the contour map of the efficiency $\eta$ in terms of $\beta_3$ and $\zeta$ in 
Fig. \ref{Compton_eff}.

Although $\beta_3$ is arbitrary as long as $\alpha_3>0$, $\zeta$ is constrained 
as $\zeta_{\rm min}<\zeta<0$ in order for the particle 2 can reach the horizon,
where
the minimum value $\zeta_{\rm min}$ depends on the spin $s_2$.
For $s_2=s_{\rm min}$, we find $\zeta_{\rm min}=-3.890$.
We then obtain the maximum efficiency for the Comptom scattering as
$\eta_{\rm max}=26.85$ in the limit of 
 $\beta_3\rightarrow -\infty$.
If $s_2=0$, the maximal efficiency is $\eta_{\rm max}\approx 13.93$, which is 
consistent with the results by Schnittman\cite{Schnittman:2014zsa}
 and Leiderschneider-Piran\cite{Leiderschneider:2015kwa}. \\[1em]

\subsection{Efficiency of inverse Compton scattering}
Since the particles 1 and 2 plunge from infinity,
we have the constraint $E_1\ge 1$ and $E_2\ge 0$.
We then assume that $E_1=1$,
 and find the maximal value of $E_3$ 
as well as the minimal value of $E_2$.
Since $E_3$ is determined only by $\alpha_3$ and $s_1$, we first discuss 
$E_3$.

\begin{figure}[h]
\includegraphics[width=6cm]{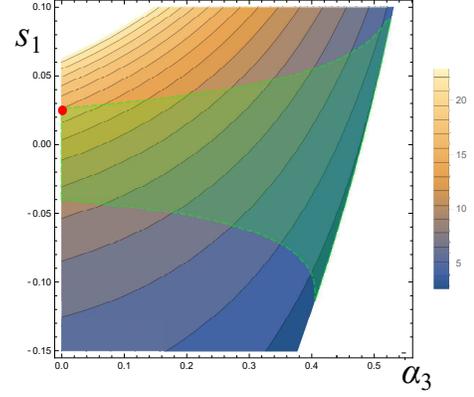}
\caption{The contour map of ${E_3}$ in terms of $\alpha_3$ and $s_1$.
The timelike condition for the particle 3 is satisfied in the light-green shaded region.
The maximum value of $E_3=E_{3,{\rm max}}=15.64$ is obtained 
at the red point $(\alpha_3, s_1)=(0, 0.02679)$.}
\label{Inverse_Compton_E3}
\end{figure}

In Fig.\ref{Inverse_Compton_E3}, 
we show the contour map of $E_3$ in terms of $\alpha_3$ and $s_1$.
The red point, which is $(\alpha_3,s_1)=(0, 0.02679)$, gives the maximal value of $E_3$.

If $E_2\rightarrow 0$ is possible, it gives the minimal value of $E_2$ and then the maximal efficiency is given by $\eta_{\rm max}=E_{3,{\rm max}}$.
Hence, assuming   $\alpha_3=0+$ and $s_1=0.02679$, 
we analyze whether $E_2\rightarrow 0$ is possible or not.
From Eq. (\ref{calP}), we find the asymptotic behavior of ${\cal P}$ as
\beann
{\cal P}\approx 
8E_3 \zeta \beta_3\left[{E_3(2+s_1)\over (1-s_1)f(s_1, E_3, 0)}-1\right]
\,,
\enann
if $\zeta\beta_3\rightarrow \infty$.
It gives $E_2\rightarrow 0$.
 $\zeta$ is constrained as $-\infty<\zeta<0$ because 
 the particle 2 is nonspinning, while 
$\beta_3$ is arbitrary as long as $\alpha_3>0$.
As a result, we obtain $E_2\rightarrow 0$ is obtained
 in the limit of $\zeta \beta_3\rightarrow \infty$. 
$\beta_3$ must be negative.
Hence, we find the maximum efficiency $\eta_{\rm max}\approx
15.64$ for the  inverse Compton scattering.
For $s_1=0$, the maximum efficiency becomes $\eta_{max}=7+4\sqrt{2}
\approx 12.66$, which is consistent with the result by Leiderschneider and Piran\cite{Leiderschneider:2015kwa}.

 \section{Concluding Remark}
 \label{Concluding_Remark}
We have analyzed the maximal efficiency of the energy extraction from 
the extreme Kerr black hole by collisional Penrose process of spinning test particles.
We summarize our result in Table \ref{summary}.

For the collision of two massive particles ({\bf MMM}+), we obtain the maximal efficiency is about $\eta_{\rm max}\approx 15.01$, which is 
more than twice as large as the case of the collision of non-spinning particles.
It happens when the particle 1 with $E_1=\mu $, $J_1=2\mu M$ and $s_1\approx0.01379\mu M$
and the particle 2 with $E_2=\mu $, $-0.5418\mu M<J_2<2\mu M$ and $s_2=s_{\rm min}\approx 0.2709\mu M$
plunge from infinity, and  collide near the horizon.
After collision, the particle 3 with $E_3\approx 30.02\mu $ and $J_3\approx 60.03\mu M$ escapes into infinity, while
the particle 4 with $E_4\approx -28.02\mu $ and $-58.57\mu M<J_4<-56.03\mu M$ falls into the black hole.

As for the collision of a massless and massive particles, we obtain the maximal efficiency $\eta_{\rm max}\approx 26.85$ for the case of {\bf PMP}+(the Compton scattering),
which is almost twice as large as the nonspinning case.
In the case of {\bf MPM}+(the inverse Compton scattering), however, we find
 $\eta_{\rm max}\approx 15.64$, which value is not so much  larger than the nonspinning case.
 It is because that the timelike condition forces the  magnitude of spin not to be
 so large for the energetic spinning particle. 

Although we have presented some examples to give a large efficiency of the energy extraction from a rotating black hole, the following cases should also be studied:
\begin{widetext}

\begin{table}[H]
\begin{center}
  \begin{tabular}{|c||c|c|c|c|}
\hline 
\raisebox{-6pt}{collisional process}&
spin &input energy& output energy  &maximal\\ [-0.4em]
&$(s_1,s_2)$&$(E_1,E_2)$&($E_3$)&efficiency\\
\hline \hline 
{\bf MMM}+& non-spinning &\raisebox{-6pt}{$(\mu,\mu)$}&$12.66 \mu$&$6.328$
\\[-.4em] \cline{2-2}\cline{4-5}
(Collision of  Two Massive Particles ) &  $(0.01379\mu M, -0.2709\mu M)$&&$30.02 \mu$ &$15.01$ \\ [.1em]
 \hline
{\bf PMP}+& non-spinning&\raisebox{-6pt}{$(+\infty,\mu)$}&$+\infty$&$13.93$
 \\[-.4em]  \cline{2-2}\cline{4-5}
(Compton Scattering) & $(0, -0.2709\mu M)$&&$+\infty$ &$26.85$\\ [.1em]
 \hline 
 {\bf MPM}+ &  non-spinning&\raisebox{-6pt}{$(\mu,0)$}&$12.66 \mu$&$12.66$
 \\[-.4em]  \cline{2-2}\cline{4-5}
(Inverse Compton Scattering) & $(0.02679\mu M, 0)$&&$15.64 \mu$ &$15.64$\\
\hline 
 \end{tabular}
    \caption{The maximal efficiencies and energies for three processes. We include 
    the nonspinning case obtained by \cite{Leiderschneider:2015kwa} as a reference. 
    The maximal efficiencies and maximal energies 
are enhanced twice or more when the spin effect is taken into account. Following \cite{Leiderschneider:2015kwa}, 
we use the symbols of {\bf MMM}+, {\bf PMP}+, {\bf MPM}+ for each process,
where + means the case of $\sigma_1=1$.
}
\label{summary}
\end{center}
\end{table}

\end{widetext}
${\bf [1]}$  \underline{Nonextreme black hole}\\
The spin of the astrophysical black hole
may not exceed $a/M=0.998$ as pointed out 
by Thorne\cite{thorne1974disk}.  Hence we should analyze the efficiency 
for a nonextreme black hole.\\[.2em]
${\bf [2]}$  \underline{Super-Penrose process }\\
We have not analyzed the case (4) : Collision of two subcritical particles.
If $\sigma_1=1$, which is not a natural initial condition for a subcritical particle, there is no upper bound for the efficiency\cite{Leiderschneider:2015kwa}. 
This super-Penrose process may be interesting to study for spinning particles too, although there still exists the question about its initial set up\cite{Zaslavskii:2015fqy}.
Recently it was discussed in \cite{Liu:2018myg}, 
but the timelike condition has not been taken into account.\\[.2em]
${\bf [3]}$ \underline{Spin transfer}\\
Since a spin plays an important role in the efficiency,
it is also interesting to discuss a transfer of spins.  
For example, $s_1=s_2=s_{\rm min}\approx -0.27$ to $s_3=0$ and 
$s_4=2s_{\rm min}\approx -0.54$.\\[.2em]
${\bf [4]}$ \underline{Collision of particles in off-equatrial-plane orbits}\\
In \cite{Leiderschneider:2015kwa}, they also analyzed the collision of the particle 
in  off-plane orbits, which gives the maximal efficiency for the case of $\sigma_1=-1$.
Although it may be interesting to analyze the orbits not in the equatorial plane,
the equations of motion for a spinning particle are not integrable.
As a result, such an analysis would be very difficult. \\[.2em]
${\bf [5]}$ \underline{Back reaction effect}\\
In this paper, we have adopted a test particle approximation. 
However because of  lack of the back reaction, 
it may not reveal the proper upper bound 
on the efficiency of the energy extraction. 
In the Reissner-Nordstr\"{o}m spacetime, 
we could perform such an analysis for the collision of charged shells
\cite{Nakao:2017xwe}.
However it would be difficult to analyze the back reaction effect in 
Kerr black hole background although it is important.

Finally one may ask how large the magnitude of spin can be  in a realistic
 astrophysical system since 
we have assumed a theoretically (or logically) allowed value of a spin in this paper.
The orbital angular momentum is given by $|\vect{L}|=|\vect{r}\times \vect{p}|\sim
R_{\rm orbit} \times \mu v\gsim O(\mu M)$, while the spin angular momentum is $s\sim 
R_{\rm body}\times \mu  v\gsim O(\mu^2)$.
Hence the ratio $s/L\sim R_{\rm body}/R_{\rm orbit}$ should be small for a test particle approximation.
In fact, if a test particle is a black hole ($s\leq \mu^2$), we find 
$s/\mu M = s/\mu^2 \times (\mu/M)\ll 1$.
Hence the value assumed here may be too large for astrophysical objects.
However, for a fast rotating star, $s$ can be much larger than $\mu^2$.
For example, we find $s/\mu^2\lsim 500$ for a fast rotator $\alpha$ LEONIS (REGULUS)
\cite{0004-637X-628-1-439}. 
Hence the validity of the test particle approximation would be marginal in this case.
The present spin effect might become important when we extend beyond a test particle 
limit including  nonlinear 
or nonperturbed process.

\section*{Acknowledgments}

We would like to thank Tomohiro Harada and Kota Ogasawara for useful discussions.
This work was supported in part by JSPS KAKENHI Grants No. JP16K05362 (KM)  
and No. JP17H06359 (KM).

~~\\

\newpage

\bibliography{spin_Penrose_0808}


\newpage

\appendix
\section{The case (3) with $\sigma_3=1$}
\label{appendix}

\subsection{\bf Case [A] MMM ({\rm Collision of two massive particles})}
In this case,  the condition $ E_3\leq E_{3, {\rm cr}}$ must be satisfied.
As a result, $E_{3,+}$, which is the larger root of Eq. (\ref{eq_E3}), is excluded.
The possible solution  is 
\beann
E_3&=&E_{3,-}:={{\cal B}-\sqrt{{\cal B}^2-{\cal A}{\cal C}} \over {\cal A}}
\,.
\enann
$E_{3, {\rm cr}}$ increases monotonically with respect to $\alpha_3$.
$E_{3, {\rm cr}}$ is positive  for
 $\alpha_3< \alpha_{3, \infty}:={2+s_2 \over 2(1+s_2)}$, 
and 
$E_{3, {\rm cr}}\rightarrow \infty$ as $\alpha_{3}\rightarrow \alpha_{3,\infty}$, while
beyond $ \alpha_{3, \infty}$, $E_{3, {\rm cr}}$ becomes negative, which case should be excluded.
As $\alpha_3$ increases, $E_3$  also increases but faster than $E_{3, {\rm cr}}$
 and reaches the upper bound $E_{3, {\rm cr}}$ at
some value of $\alpha_3=\alpha_{3,{\rm cr}}$.
 
\begin{widetext}

\begin{figure}[h]
\includegraphics[width=15cm]{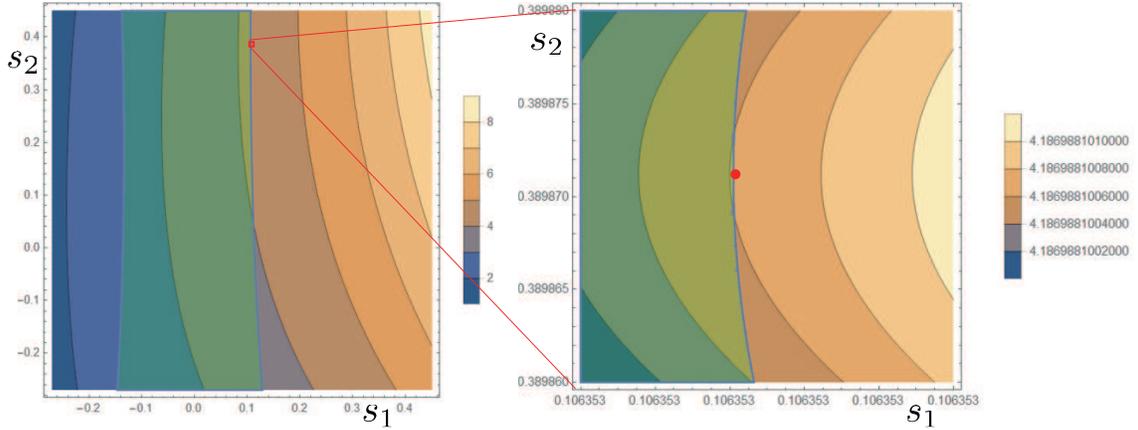}
\caption{The contour map of ${E_{3,{\rm cr}}} (s_1,s_2)$ 
with $\alpha_3=\alpha_{3,{\rm cr}}$, which gives the largest value of $E_3$
for given $s_1$ and $s_2$. The timelike condition for the particle 3 is satisfied in the light-green shaded region.
The right figure is enlarged near the maximal point ($(s_1,s_2)\in (0.1063534539, 
0.1063534540)\times (0.38986, 0.38988)$).
The maximal value $E_{3, {\rm max}} \approx 4.187$ is obtained at the red point 
 $(s_1, s_2) \approx (0.10635, 0.3899)$ with $\alpha_3\approx 
0.5342$. 
}
\label{Appendix_E3}
\end{figure}

\end{widetext}

For given 
values of $s_1$ and $s_2$, we find 
the quadratic equation for $\alpha_{3,{\rm  cr}}$
  from the condition  $E_{3, {\rm cr}} = E_3$.
Inserting the solution $\alpha_{3,{\rm  cr}}(s_1, s_2)$ into the definition of $E_3$, 
we find the largest value of $E_3$, $E_{3, {\rm cr}} (s_1,s_2)$ in terms of $s_1$ and $s_2$.
We show  its contour map in Fig.\ref{Appendix_E3},
in which  we also plot the timelike condition by the light-green shaded region.
This gives
 the maximal value $E_{3, {\rm max}} \approx 4.187$ 
at the red point $(s_1, s_2) \approx (0.10635, 0.3899, 0.534)$ with $\alpha_3\approx 
0.5342$.

\begin{figure}[h]
\includegraphics[width=3.9cm]{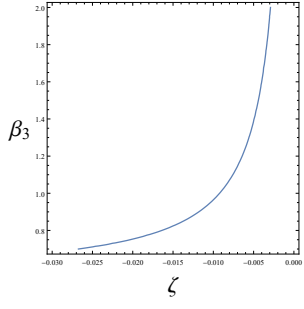}
\caption{The relation between $\zeta$ and $\beta_3$ for ${E_2}=1$
when $E_3$ takes  the maximal value.
}
\label{Appendix_E2}
\end{figure}

We then check that $E_2=1$ is possible for the  above parameters.
Fig.\ref{Appendix_E2} shows the relation between $\zeta$ and $\beta_3$ for $E_2=1$.
We conclude that the maximal efficiency is $\eta_{\rm max}=E_{3, {\rm max}}/2 \approx 
2.093$.

\subsection{{\bf Case [B] PMP} {\rm Compton scattering}}
\begin{figure}[h]
\includegraphics[width=6cm]{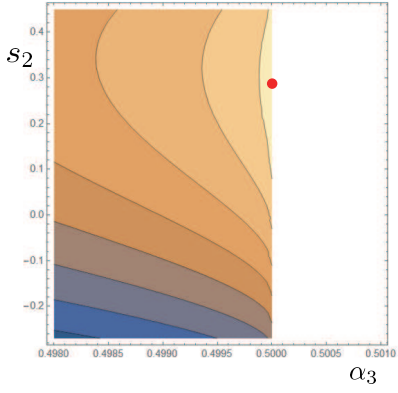}
\caption{The  contour map of the magnification factor ${\cal S}$ in terms of $\alpha_3$ and $s_2$ for  the Compton scattering in the case of $\sigma_3=1$.
The maximal value ${\cal S}_{\rm max}\approx 3.876$ is found at  the red point $(\alpha_3, s_2)\approx (0.5, 0.2887)$.
} 
\label{Compton_mag}
\end{figure}

In this case, since the particle 1 is massless, we first draw the contour map of the magnification factor {\cal S}, which is defined by Eq. (\ref{Compton_magnification}), in Fig. \ref{Compton_mag}.
The maximal value of ${\cal S}$ is ${\cal S}_{\rm max}\approx 3.876$, which  is found at  the red point $(\alpha_3, s_2)\approx (0.5, 0.2887)$.
We note that the value of $\alpha_3$ must be either $\alpha_3\leq 0.5$ or $\alpha_3\geq 1.5$ in order to find a real value of ${\cal S}$.
However there is the constraint as $\alpha_3\leq 1$ from the 
future-directed proper time condition of $dt/d\lambda>0$, although the larger value of ${\cal S}$ is possible for $\alpha_3>1.5$.
As a result, we find the above maximal value of ${\cal S}_{\rm max}$.

\begin{figure}[h]
\includegraphics[width=6cm]{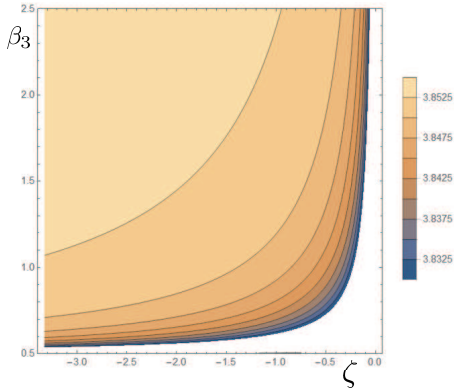}
\caption{The contour map of the efficiency $\eta$  in terms of $\zeta$ and $\beta_3$.
The maximal value of the efficiency is about 3.853 in the limit of $\zeta\rightarrow \zeta_{\rm min}\approx 
-3.326$ and $\beta_3\rightarrow \infty$. } 
\label{Compton_eff2}
\end{figure}

Fig. \ref{Compton_eff2} show that the efficiency is more than 3.85 in the wide range of 
parameters ($\zeta_{\rm min}\approx -3.326 \lsim \zeta\lsim -0.5$ and $\beta_3\gsim 0.5$) and it takes the maximal value 3.854,
which is close to ${\cal S}_{\rm max}$, in the limit of $\zeta\rightarrow \zeta_{\rm min}\approx 
-3.326$ and $\beta_3\rightarrow \infty$.

\begin{figure}[h]
\includegraphics[width=6cm]{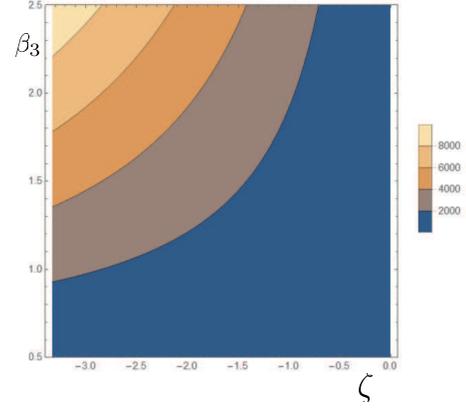}
\caption{The  contour map of $E_1/E_2$ in terms of $\zeta$ and $\beta_3$.
 The dynamic range of $E_1/E_2$  is very wide from 1 to $10^4$.
} 
\label{Compton_E1_E2}
\end{figure}

However we find  $E_2=0$ when $\alpha_3=0.5$.
Since  $E_2\geq 1$, we cannot choose $\alpha_3=0.5$.
Hence choosing $\alpha_3=0.49999$ as well as $s_2=0.288675$, 
we show the contour maps of the efficiency $\eta$ and $E_1/E_2$ in Fig. \ref{Compton_eff2} and Fig. \ref{Compton_E1_E2}.

The value of $E_1/E_2$ is larger than 1000 in the above range of parameters.
Since $E_2\geq 1$, we find the efficiency is about  3.85 when $E_1\gsim 10^3$
(see Fig. \ref{Compton_E1_E2}). 
If we take the limit of $\alpha_3=0.5$, the maximal efficiency is obtained 
$\eta_{\rm max}={\cal S}_{\rm max}\approx 3.876$ when $E_1\rightarrow \infty$.

The above result shows the efficiency can be larger than 3.5 but never exceeds  ${\cal S}_{\rm max}\approx 3.876$
when  the plunged photon energy is much larger than the particle 2 rest mass.

Note that we find the upper bound of the efficiency is about 3.732
 even for the nonspinning case ($s_2=0$).


\subsection{{\bf Case [C] MPM} {\rm Inverse Compton scattering}}
\begin{figure}[h]
\includegraphics[width=6cm]{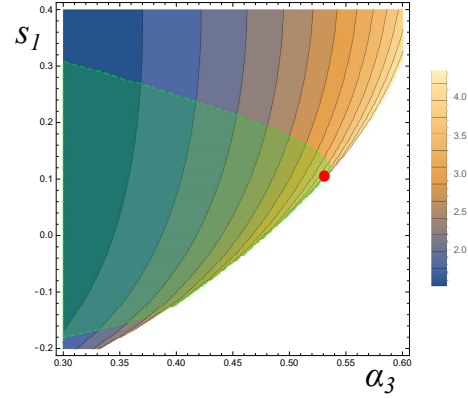}
\caption{The contour map of $E_3$ in terms of $\alpha_3$ and $s_1$ for the inverse Compton scattering in the case of $\sigma_3=1$. The timelike condition for the particle 3 is satisfied in the green shaded region.
The maximal value ${E}_{3}\approx 4.202$ is found at  the red point $(\alpha_3, s_1)\approx (0.5331, 0.1059)$.
}
\label{Appendix_Inverse_Compton_E3}
\end{figure}

We first depict the contour map
of $E_3$ in terms of $\alpha_3$ and $s_1$ in Fig. \ref{Appendix_Inverse_Compton_E3}.

The maximal value ${E}_{3{(\rm max)}}\approx 4.202$ is obtained  
at  the red point $(\alpha_3, s_1)\approx (0.5331, 0.1059)$. 
From Eq. \ref{calP}, we find the asymptotic behavior of ${\cal P}$ as
\beann
{\cal P}\approx 
-8E_3 \zeta \beta_3\left[{E_3 g_1(s_1, \alpha_3) \over (1-s_1)f(s_1, E_3, \alpha_3)}+1\right]
\,,
\enann
when we take a limit of  $\zeta \beta_3\rightarrow -\infty$.
Since $g_1(s_1, \alpha_3)>0$ in the plotted region of Fig.\ref{Appendix_Inverse_Compton_E3},
we find $E_2\rightarrow 0$ as $\zeta \beta_3\rightarrow -\infty$. 
For the nonspinning particle 2, only the condition $ \zeta < 0$ is required. 
As a result $\beta_3$ must be positive to get $E_2\rightarrow 0$. 
We obtain  the maximal efficiency as $\eta_{\rm max}
={E}_{3{(\rm max)}}\approx 4.202$.

\end{document}